\documentclass[prb,10pt,twocolumn]{revtex4}

\usepackage{graphicx,amssymb,amsmath}

\begin{document}

\author{Jared H. Strait}
\email{jhs295@cornell.edu}
\author{Haining Wang}
\author{Shriram Shivaraman}
\author{Virgil Shields}
\author{Michael Spencer}
\author{Farhan Rana}
\affiliation{School of Electrical and Computer Engineering, Cornell University, Ithaca, NY, USA}

\title{Very Slow Cooling Dynamics of Photoexcited Carriers in Graphene Observed by Optical-Pump Terahertz-Probe Spectroscopy}

\begin{abstract}
Using optical-pump terahertz-probe spectroscopy, we study the relaxation dynamics of photoexcited carriers in graphene at different temperatures. We find that at lower temperatures the tail of the relaxation transients as measured by the differential probe transmission becomes slower, extending beyond several hundred picoseconds at temperatures below 50K. We interpret the observed relaxation transients as resulting from the cooling of the photoexcited carriers via phonon emission. The slow cooling of the photoexcited carriers at low temperatures is attributed to the bulk of the electron and hole energy distributions moving close enough to the Dirac point that both intraband and interband scattering of carriers via optical phonon emission becomes inefficient for removing heat from the carriers.  Our model, which includes intraband carrier scattering and interband carrier recombination and generation, agrees very well with the experimental observations.  
\end{abstract}

\maketitle

In recent years, graphene has gathered much interest for electrical and optical applications due to its unusual band structure and properties \cite{Geim07, Abergel10}.  Graphene has been involved in an array of applications that highlight its versatility and novelty as a platform for electronic, plasmonic, optical/IR, and terahertz devices \cite{Bonaccorso10, BaoSatAbs09,XiaPhotodet09,RanaGTPO08,Song11,Shepard08}.  Realization of many of the graphene based devices relies on a good understanding of the nonequilibrium carrier and phonon processes and their associated time scales. In particular, the dynamics associated with the cooling and recombination of photoexcited carriers are of interest in demonstrated and proposed graphene optoelectronic and terahertz devices \cite{XiaPhotodet09, RanaGTPO08, Song11}. 

Relaxation dynamics of photoexcited carriers in graphene have recently been studied using ultrafast optical/IR pump-probe spectroscopy by several groups including the authors \cite{LuiPhotoLum10,DawlatyUltrafast08,WangPhonon10,Sun08,Huang10,Newson09,Malard11,KangOPLife10}. These measurements, which typically measure the relaxation of the high energy tail of the carrier distribution, have shown that the photoexcited carriers thermalize within few tens of femtoseconds to generate a hot carrier distribution.  This hot distribution then cools rapidly via optical phonon emission on a time scale of hundreds of femtoseconds.  Within one picosecond, the carrier and the optical phonon temperatures equilibrate, and carrier cooling slows.  At this point, cooling is limited by the exchange of energy between the carriers and the optical phonons and the subsequent anharmonic decay of optical phonons into acoustic phonons. The questions that still remain unanswered pertain to the nature of the relaxation dynamics over much longer time scales and to the role played by carrier generation and recombination processes in the observed relaxation dynamics.  The answer to these questions is interesting both from the perspective of practical devices and also from a theoretical point of view.  For example, theoretical groups have recently pointed out that the cooling of hot carriers in doped graphene is very slow when a majority of the carrier distribution is below the optical phonon energy.  In this case, carrier cooling can occur only via acoustic phonon emission \cite{Tse09,Bistritzer09}. Compared to other semiconductors, graphene stands out due to its rather large optical phonon energies ($\sim$ 0.196 eV and 0.162 eV).  Therefore, the optical phonon energy bottleneck in carrier cooling is expected to play an important role in many graphene-based electronic and optical devices \cite{XiaPhotodet09,Song11,Shepard08}. In this paper, we report observations of this bottleneck in the cooling of photoexcited carriers for the first time.   
  
Ultrafast terahertz spectroscopy is a useful tool to study the relaxation dynamics of the low energy carriers near the Dirac point. Previous studies of graphene using optical-pump terahertz-probe spectroscopy, carried out at room temperature by the authors \cite{Paul08} and others \cite{Choi09}, attributed the observed relaxation transient occurring over a 1-10 ps time scale to carrier recombination.  However, recent theoretical results reported by the authors and others show that the interband recombination and generation mechanisms in graphene, such as Auger scattering and impact ionization \cite{RanaAuger07,WinzerAuger10}, optical phonon scattering \cite{RanaPhonon09}, and plasmon scattering \cite{RanaPlasmon10}, can have characteristic times much shorter than one picosecond. In particular, plasmon scattering can be extremely fast, with time scales on the order of a few hundred femtoseconds  \cite{RanaPlasmon10}.  In this paper, we present results from optical-pump terahertz-probe spectroscopy of photoexcited carriers in graphene at different temperatures.  We vary the temperature in order to better understand the role played by several scattering processes in the observed relaxation transients.  Our results show that the tails of the relaxation transients, as measured by the differential probe transmission, become remarkably slow at low temperatures.  They extend well beyond several hundred picoseconds at temperatures below 50K. 

We also present a theoretical model to explain the measured data.  Our model includes intraband carrier scattering from optical and acoustic phonons as well as interband carrier recombination and generation from optical phonons, plasmons, Auger scattering, and impact ionization. Our model shows that the photoexcited electrons and holes equilibrate with each other within one picosecond due to the very fast recombination and generation processes.  This fast equilibration causes the electron and hole Fermi levels to merge.  The experimentally observed relaxation transients beyond a few picoseconds are then due entirely to the cooling of the carriers.  We attribute the very slow tails of the relaxation transients observed at low temperatures to the bulk of the carrier energy distributions moving close enough to the Dirac point such that both intraband and interband scattering of carriers via optical phonon emission become inefficient.  For completely undoped samples, with symmetric electron and holes distributions, this occurs when the bulk of the electron and hole distributions move below half the optical phonon energy.  Our results compliment the earlier theoretical predictions \cite{Tse09,Bistritzer09}, and our model agrees very well with our measurements at all temperatures. 
                                   
Graphene samples used in this work were grown epitaxially via thermal decomposition of the SI-SiC ($000\bar{1}$) surface \cite{Berger06}.  The samples were characterized with Raman and optical transmission spectroscopy to estimate the number of graphene layers. In experiments, the samples were placed in a Helium cryostat. Optical pump excitations were performed with $\sim$90 fs optical pulses, with 780 nm center wavelength, obtained from a 81 MHz Ti:Sapphire laser.  The optical pump pulses had maximum energies of $\sim$11.4 nJ and were incident on the sample from an angle with a spot size on the sample of $\sim$1.0 mm$^{2}$.  The maximum estimated photoexcited carrier density was $\sim5\times10^{10}$ 1/cm$^{2}$. The non-equilibrium carrier distribution in the graphene layers was probed by monitoring the differential transmission of few-cycle terahertz pulses, with a peak frequency of $\sim$1 terahertz, focused to a 0.5 mm$^{2}$ spot size on the sample.  The terahertz pulses were generated and detected in a terahertz time-domain spectrometer setup based on photoconductive antennas \cite{Katzenellenbogen91} and had a power SNR > $10^6$. The temporal resolution of our pump-probe measurement was limited not by the duration of the terahertz pulses but by the duration of the optical pulses to $\sim$1 ps, considering the fact that the optical pump was incident on the sample from an angle. The optical pump and terahertz probe beams were chopped at 333 Hz and 400 Hz, respectively, and the differential transmission signal was detected using a lock-in amplifier at the sum frequency.

\begin{figure}[tbp]
	\centering
		\includegraphics[width=.40\textwidth]{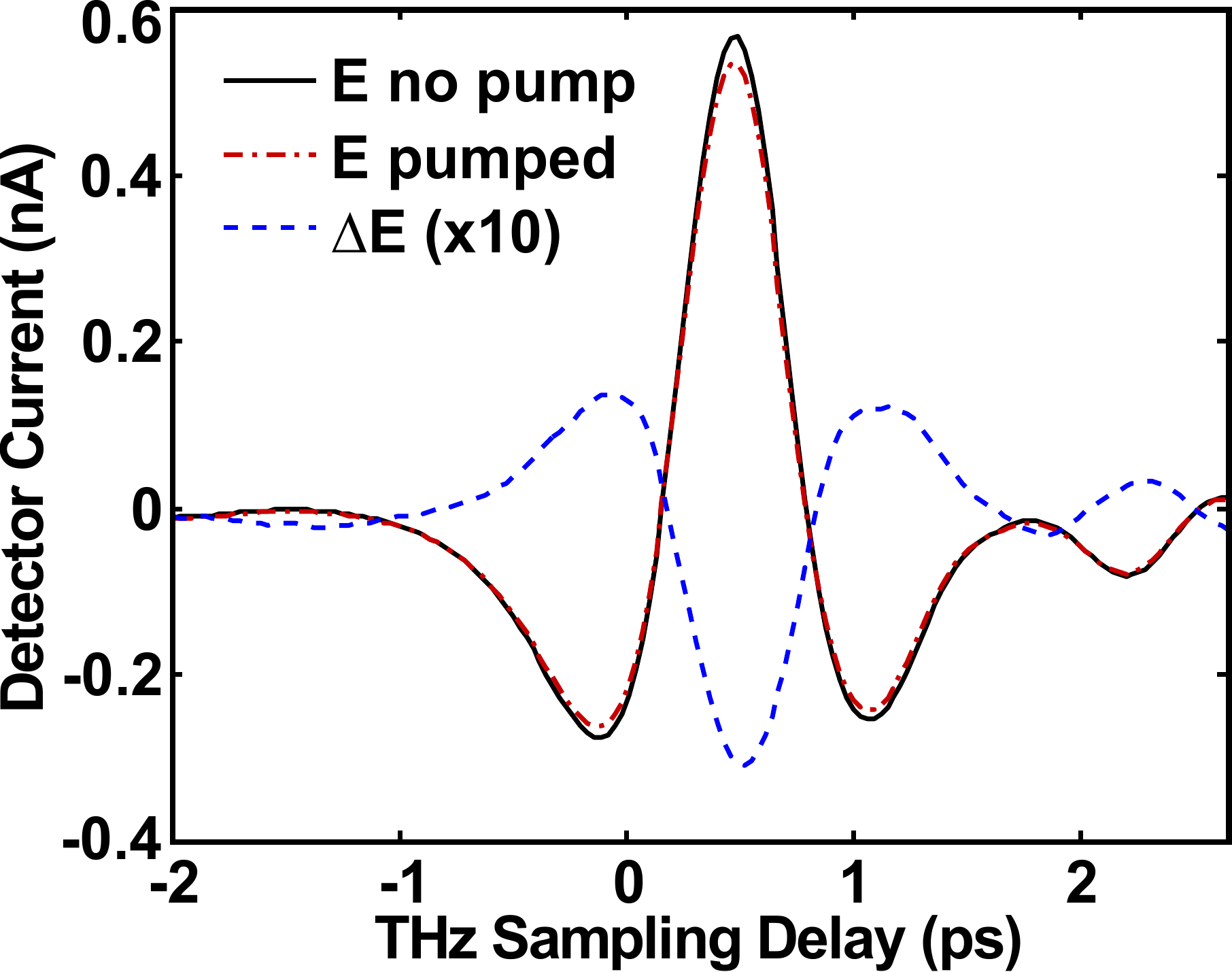}
	\caption{The electric fields of terahertz pulses transmitted through epitaxial graphene at $T_{sub}=18$K are plotted with and without optical pumping. Also shown is $\Delta E = (E{\rm\ pumped})-(E{\rm\ no\ pump})$.  A small phase shift of less than 50 fs between the two pulses indicates that the Drude momentum scattering time, $\tau$, is small.}
	\label{fig:deltaE}
\end{figure}

The frequency dependence of the graphene conductivity has been shown to have a Drude like behavior \cite{Dawlaty08}. Changes in a terahertz probe pulse from the graphene layers are in general complex.  However, in cases where the carrier momentum scattering time $\tau$ is much shorter than the time scale associated with all other dynamics of interest, then the transmitted terahertz pulse is essentially the input terahertz pulse with an amplitude modulation given by the instantaneous carrier distribution \cite{Paul08}. Fig.\ref{fig:deltaE} shows the field amplitude $E$ of a terahertz pulse transmitted through a 14 mono-layer (ML) graphene sample with and without optical pumping.  Their difference, $\Delta E$, is also shown. The maximum observed phase shift is less than 50 fs, indicating that the electronic scattering time $\tau$ is indeed very short in our samples. Therefore, we track only the peak amplitude of the terahertz probe pulse as a function of the probe delay.

\begin{figure}[tbp]
	\centering
		\includegraphics[width=.40\textwidth]{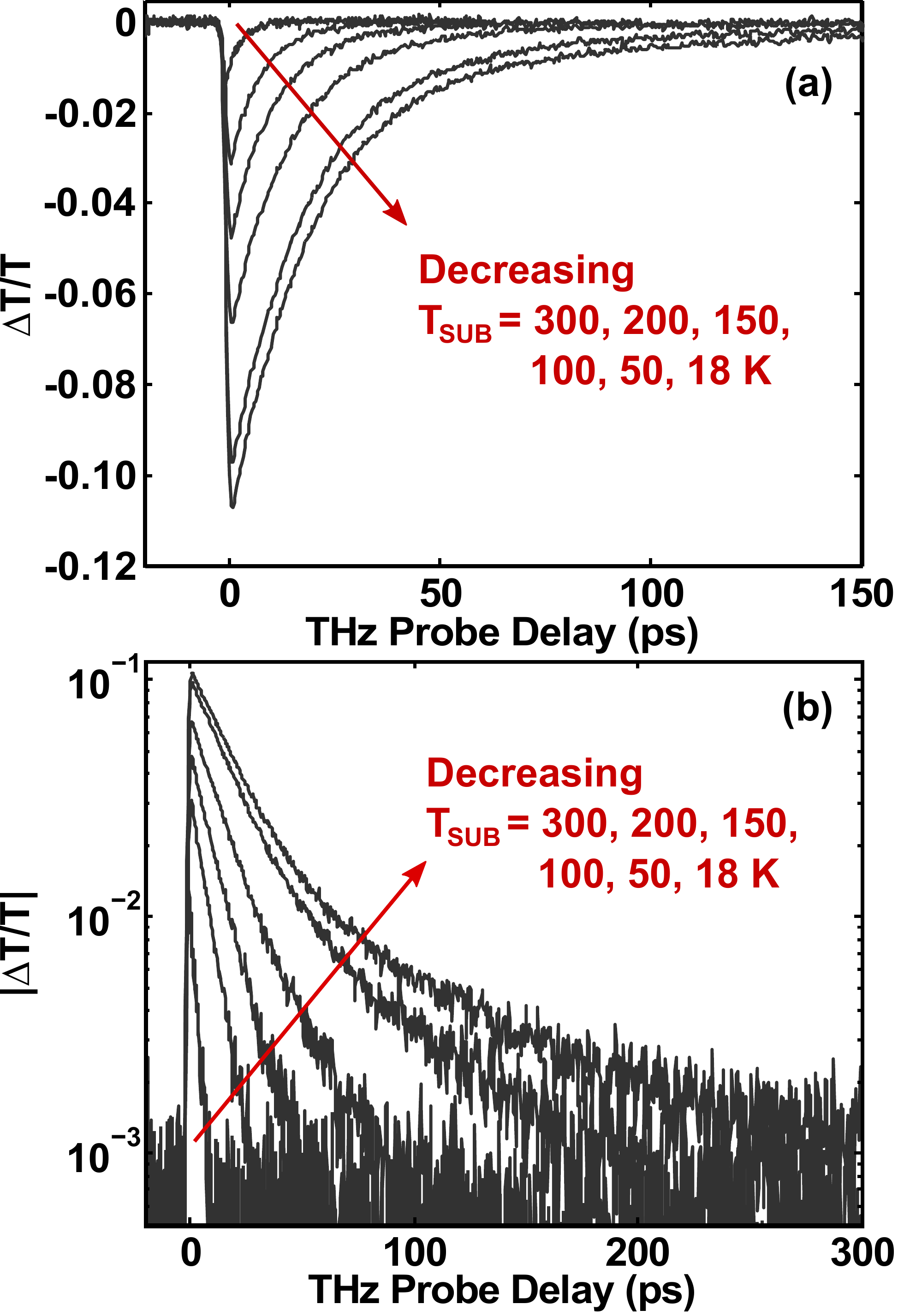}
	\caption{(a) Measured differential terahertz probe transmission $\Delta T/T$ is plotted as a function of the probe delay for a $\sim$14 layer epitaxial graphene sample at six different substrate temperatures ($T_{sub}$=300, 200, 150, 100, 50, and 18K).  The estimated photoexcited carrier density is $5\times10^{10}$ 1/cm$^2$. Lower substrate temperatures result in larger peak $|\Delta T/T|$ values and slower relaxation rates. (b) The same data as in (a) shown on a log scale.}
	\label{fig:vsTemp}
\end{figure}

The normalized differential transmission amplitude $\Delta T/T$ of the terahertz probe pulse, as a function of the probe delay time, is shown in Fig.\ref{fig:vsTemp} for different substrate temperatures for the same 14 ML graphene sample. The following observations can be made from this data. Immediately after photoexcitation, $\Delta T/T$ decreases on a time scale of $\sim$1 ps, although this observation is limited by the resolution of our setup.  After $\sim$1 ps, $\Delta T/T$ recovers over time scales that strongly depend on the substrate temperature $T_{sub}$.  Decreasing the substrate temperature causes the recovery times to increase from tens of picoseconds at room temperature to hundreds of picoseconds at low temperatures.  At low substrate temperatures, two distinct time scales are observed; a fast recovery phase lasting to about 50 ps and a much slower phase lasting to hundreds of picoseconds. Measurement of the transients over time scales longer than 300 ps were not possible with our setup. Also, the peak magnitude $|\Delta T/T|$ is larger at lower substrate temperatures, increasing by as much as an order of magnitude at $T_{sub}=18$K compared to $T_{sub}=300$K.

In this section, we discuss our model, which is in good agreement with the measured data. We assume that immediately after photoexcitation, the photoexcited electrons and holes thermalize among themselves as well as with the existing carriers through carrier-carrier scattering on a time scale of tens of femtoseconds \cite{Elsaesser09}.  After thermalization, the carriers have Fermi-Dirac energy distributions with well defined Fermi-levels ($E_{fe}$ and $E_{fh}$ for electrons and holes respectively) and a common temperature $T_{eh}$.  We simulate the subsequent evolution of the electron, hole, and phonon distributions using rate equations for the electron density $n_{e}$, hole density $n_{h}$, electron/hole temperature $T_{eh}$, optical phonon occupation numbers $n_{\Gamma}$ and $n_{K}$ at the $\Gamma$ and $K$-points of the Brillouin zone, and the acoustic phonon temperature $T_{a}$.  The intraband and interband electron and hole scattering via optical phonons is described according to the models presented by Wang et.\ al.\ \cite{WangPhonon10} and Rana et.\ al.\ \cite{RanaPhonon09}.  The interband electron and hole scattering via Auger and impact ionization is described using the model of Rana et.\ al. \cite{RanaAuger07}. The intraband electron and hole scattering via longitudinal acoustic phonons is described according to the model presented by Suzuura and Ando \cite{Ando02} using a value of 19 eV for the deformation potential. The interband electron and hole scattering via plasmons is described using the model of Rana et.\ al. \cite{RanaPlasmon10}. The anharmonic decay of optical phonons into acoustic phonons is described phenomenologically with the time constant $\tau_{opt}$, which has typical values in the 0.5-2.5 ps range \cite{WangPhonon10,KangOPLife10,Heinz11,Mauri07}. The loss of heat from the graphene acoustic phonons into the substrate is also described phenomenologically with the time constant $\tau_{sub}$, which has values in the 25-200 ps range \cite{Mak10}.

A feature of these rate equations is that the rate of carrier temperature change, $dT_{eh}/dt$, is related to the rate of change of the carrier energy density, $dU_{e}/dt + dU_{h}/dt$, and carrier densities, $dn_{e}/dt$ and $dn_{h}/dt$, via the relation
\begin{equation}
\frac{dT_{eh}}{dt} = \frac{1}{C_{e} + C_{h}} \left( \frac{dU_{e}}{dt}  + \frac{dU_{h}}{dt} - \chi_{e}\frac{dn_{e}}{dt} - \chi_{h} \frac{dn_{h}}{dt} \right) \label{eq:1}
\end{equation}    
Here, $C_{e}$ and $C_{h}$ are the electron and hole heat capacities, and $\chi_{e}$ is given in terms of integrals of the Fermi-Dirac distribution $f(.)$ as
\begin{equation}
\chi_{e} = \frac{ \int_{0}^{\infty} dE \, E^{2} f(E_{fe},T_{eh}) (1 -  f(E_{fe},T_{eh}))}{ \int_{0}^{\infty} dE \, E f(E_{fe},T_{eh}) (1 -  f(E_{fe},T_{eh})) } \label{eq:2} 
\end{equation}
The expression for $\chi_{h}$ is obtained by substituting $-E_{fh}$ for $E_{fe}$ in the above equation. Eq.\ref{eq:1} and Eq.\ref{eq:2} show that recombination (generation) due to carrier-carrier interactions (Auger scattering, impact ionization, and plasmon emission and absorption), in which the total energy of the electrons and holes does not change, always results in an increase (decrease) in the temperature of the carriers. Also note that for all the recombination and generation models we consider, when $E_{fh}>E_{fe}$, the generation rate exceeds the recombination rate.  When $E_{fh}<E_{fe}$, the opposite is true \cite{RanaPhonon09,RanaAuger07,RanaPlasmon10}.        

\begin{figure}[tbp]
	\centering
		\includegraphics[width=.40\textwidth]{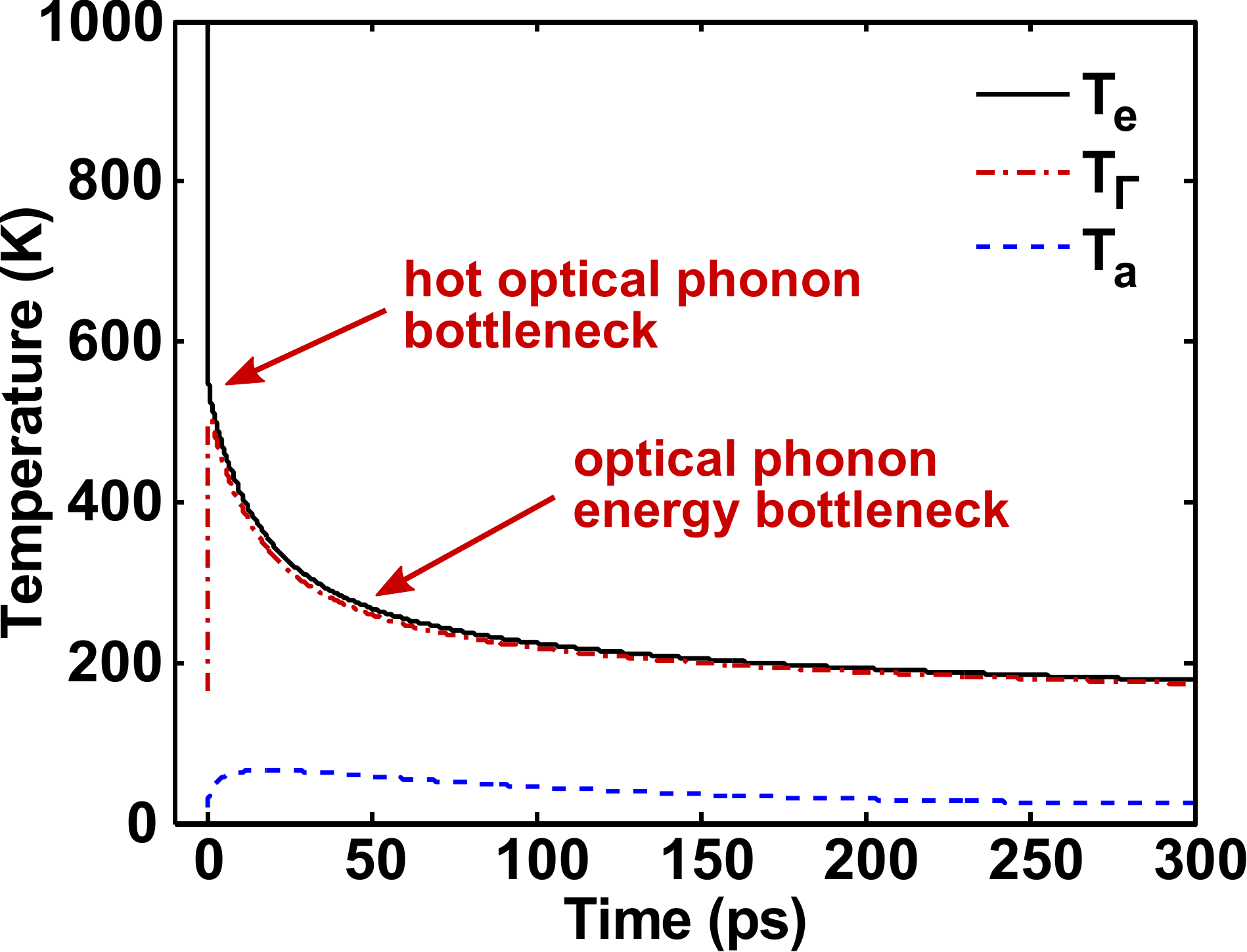}
	\caption{Simulation results of the electronic temperature, $T_{eh}$, the optical phonon temperature, $T_{\Gamma} (\approx T_{K})$, and the acoustic phonon temperature, $T_{a}$, as a function of time after photoexcitation.  $T_{sub}=18$K.}
	\label{fig:temps}
\end{figure}

\begin{figure}[tbp]
	\centering
		\includegraphics[width=.40\textwidth]{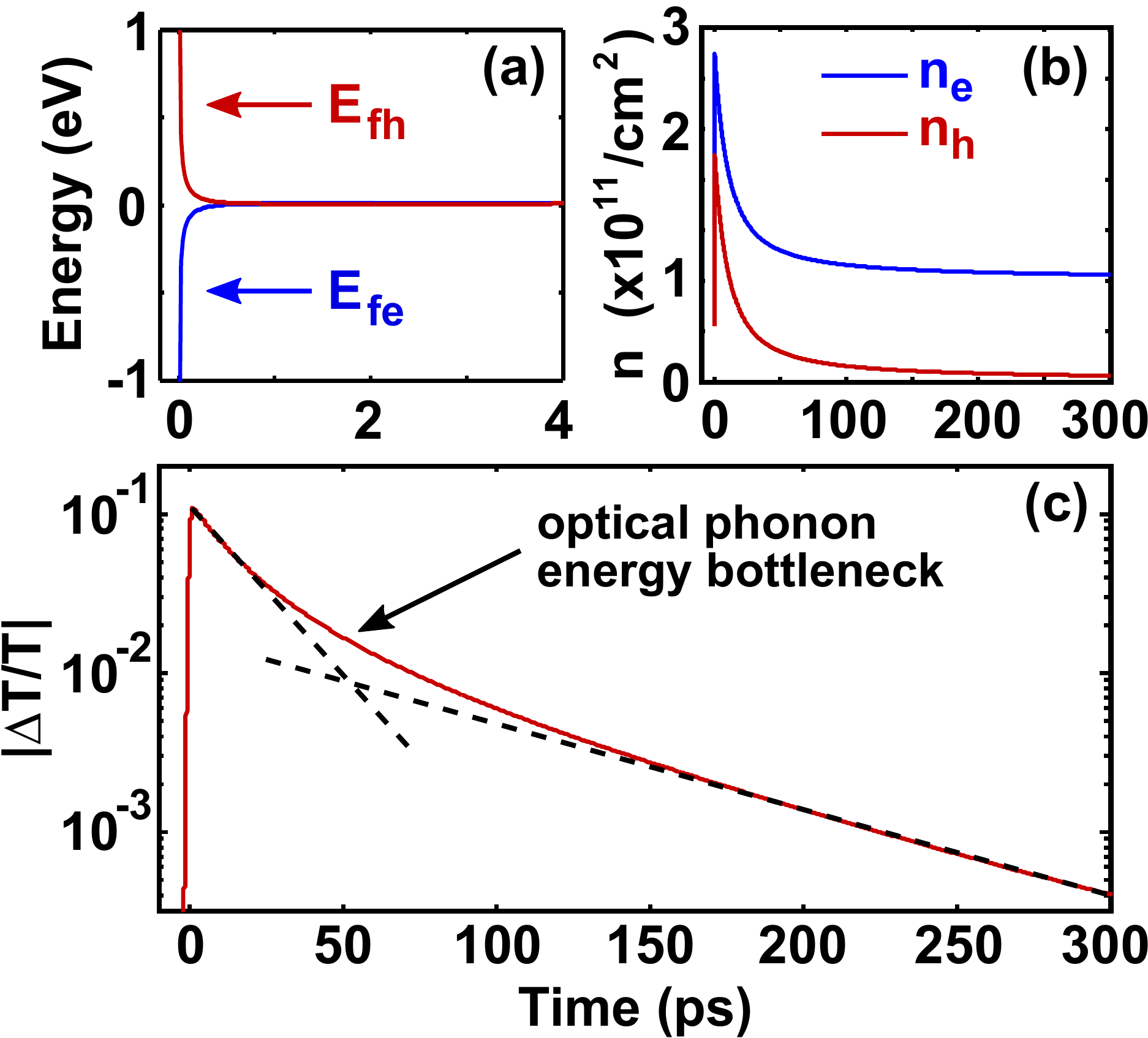}
	\caption{(a) Simulation results of the Fermi levels, $E_{fe}$ and $E_{fh}$, for electrons and holes, respectively, as a function of time after photoexcitation. (b) Simulation results for the electron and hole densities, $n_e$ and $n_h$, as a function of time. (c)  Simulation results of the differential terahertz transmission $|\Delta T/T|$ transient.  Dashed lines indicate the approximate rate of relaxation at short and long time scales.}
	\label{fig:sim}
\end{figure}

Representative results from the simulations are plotted in Fig.\ref{fig:temps} and Fig.\ref{fig:sim}. Fig.\ref{fig:temps} shows the carrier temperature $T_{eh}$, optical phonon temperatures $T_{\Gamma}(\approx T_{K})$, and the acoustic phonon temperature $T_{a}$ plotted as a function of time. Fig.\ref{fig:sim}(a) shows the electron and hole Fermi levels, $E_{fe}$ and $E_{fh}$, and Fig.\ref{fig:sim}(b) shows the electron and hole densities, $n_{e}$ and $n_{h}$. In simulations, the photoexcited carrier density was assumed to be $5\times10^{10}$ 1/cm$^{2}$, $T_{sub}=18$K, $\tau=28$ fs, $\tau_{opt}=1.6$ ps, $\tau_{sub}=100$ ps, and the sample was assumed to be slightly n-doped with doping density $N_{d}=1\times10^{11}$ 1/cm$^{2}$.  Immediately after thermalization, the carrier temperature is large ($\sim2300$K in Fig.\ref{fig:temps}) and $E_{fh} >> E_{fe}$.  In the initial phase lasting to about 0.5-1.0 ps, the carriers lose energy via optical phonon emission and their distribution cools down relatively fast.  In this phase, $E_{fh}$ exceeds $E_{fe}$, so carrier generation rate exceeds carrier recombination rate and the electron/hole densities increase. As discussed earlier, carrier generation via carrier-carrier interactions also contributes to a decrease of the carrier temperature.  Within a picosecond, the electron and hole populations equilibrate and their Fermi levels merge.  Beyond $\sim$1 ps, the electron and hole distributions can be approximately described by a common Fermi level; but strictly speaking now $E_{fe} > E_{fh}$, so recombination exceeds generation and the electron/hole densities decrease with time. In the same time frame of 0.5-1.0 ps, optical phonon emission leads to an increase in the optical phonon temperatures until the carrier and the optical phonon temperatures become nearly identical.  After this point, hot optical phonons become the main bottleneck for further carrier cooling and carrier cooling slows dramatically \cite{WangPhonon10}, as indicated in Fig.\ref{fig:temps}.  As the electron and hole distributions cool further, at some point the bulk of these distributions moves close enough to the Dirac point such that both the intraband and the interband scattering of carriers via optical phonon emission become inefficient in cooling the carriers.  For the values considered in the simulations presented in Fig.\ref{fig:temps} and Fig.\ref{fig:sim}, this occurs when the carrier temperature falls below $\sim$250K.  Since longitudinal acoustic phonon scattering is also inefficient in cooling the carriers \cite{Tse09,Bistritzer09}, the carrier cooling rate slows down further. This optical photon energy bottleneck is indicated in Fig.\ref{fig:temps}.  Note that the larger heat capacity of the acoustic phonons results in the maximum change in the acoustic phonon temperature being smaller than the maximum change in the optical phonon temperature.

The transmission of a terahertz pulse through $N$ graphene layers on a SiC substrate normalized to the transmission through the SiC substrate is given by the expression \cite{Dawlaty08},
\begin{equation}
\frac{T}{T_{\rm SiC}}=\frac{1}{1+N\sigma(\omega)\eta_0/(1+n_{\rm SiC})}
\end{equation}
Here, $n_{\rm SiC}$ is the refractive index of the SiC substrate and $\sigma(\omega)$ is the intraband conductivity of graphene given by \cite{Dawlaty08},
\begin{equation}
\sigma(\omega) = i\frac{e^2/\pi\hbar^2}{\omega+i/\tau}\int_0^\infty \!\! \left( f(E_{fe},T_{eh}) + f(-E_{fh},T_{eh}) \right) dE \label{eq:cond}
\end{equation}
Fig.\ref{fig:sim}(c) shows the calculated relative differential transmission $\Delta T/T$ of the terahertz probe pulse as a function of the probe delay.  In the first $\sim$1 ps of the simulation, the graphene conductivity increases and the terahertz transmission decreases.  This increase of conductivity is due to two factors.  First, as the temperature of the carriers decreases, the graphene conductivity as given by Eq.\ref{eq:cond} increases. This results from graphene conductivity depending on both the total number of carriers and also on the carrier distribution in energy.  For the same number of carriers, the conductivity is larger if the carrier temperature is smaller.  Second, carrier generation also contributes to an increase in the number of carriers and, therefore, an increase in the conductivity.  Beyond $\sim1$ ps, the conductivity decreases and the terahertz transmission increases.  This decrease in conductivity is due to the decrease in the carrier densities as the carriers cool down.  The relaxation of the $\Delta T/T$ transient exhibits two distinct time scales: a first fast relaxation phase lasting to about 50 ps during which both intraband and interband optical phonon emission is efficient in cooling the carrier distributions, and the second slow phase lasting longer than hundreds of picoseconds during which optical phonon emission is inefficient in cooling the carriers.

\begin{table}
\centering
\begin{tabular}{|l|l|l|l|} 
\hline
\multicolumn{4}{|c|}{Table.1: Fitting Parameters}\\ 
\hline
 & $T_{sub}$=300\ K &   $T_{sub}$=150\ K &  $T_{sub}$=18\ K \\
\hline
$N_{d}$ (1/cm$^2$) & $2.5\times 10^{11}$ & $1.4\times 10^{11}$ & $1.0\times 10^{11}$ \\
\hline
$\tau$ (fs) & 5 & 13.7 & 28.3 \\
\hline
$\tau_{opt}$ (ps) & 0.8 & 1.1 & 1.6 \\
\hline 
\end{tabular}
\end{table}

\begin{figure}[tbp]
	\centering
		\includegraphics[width=.40\textwidth]{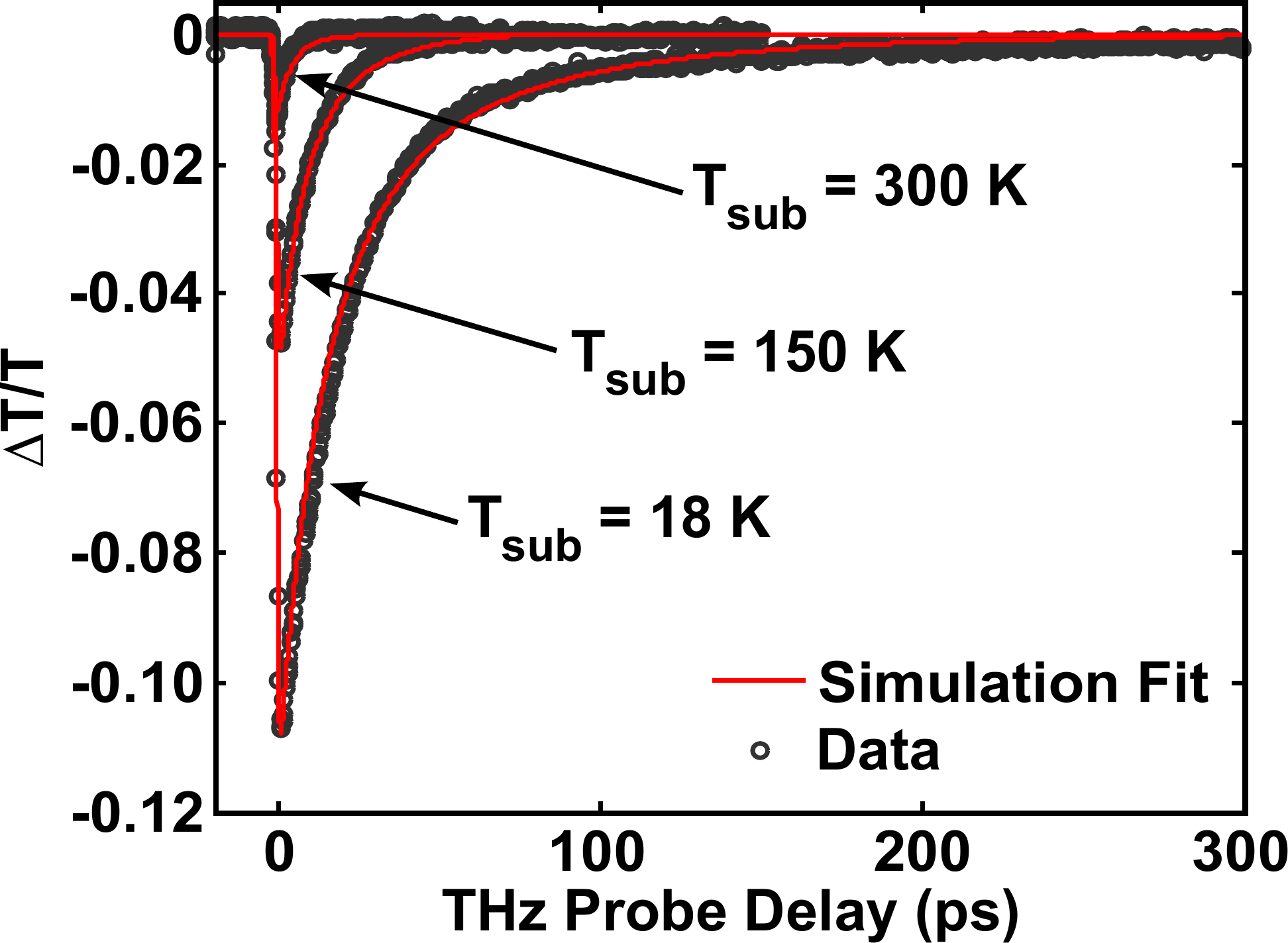}
	\caption{Measured $\Delta T/T$ transients are plotted along with the simulation fits for different substrate temperatures.}
	\label{fig:fitdata}
\end{figure}

Fig.\ref{fig:fitdata} shows the comparison between the theoretical model (solid lines) and the measurements (circles) of the differential terahertz transmission $\Delta T/T$ for three different substrate temperatures, $T_{sub}$ = 300, 150, 18K. The only three fitting parameters used in the simulations were the graphene doping density $N_{d}$, the carrier momentum scattering time $\tau$, and the optical phonon decay time $\tau_{opt}$. Table.1 gives the fitting values used for different substrate temperatures.  The time constant $\tau_{sub}$, which describes the loss of heat from the graphene layers into the substrate, was varied between 25 ps and 100 ps \cite{Mak10}; however, its value was found to have no significant effect on the simulation results for $\Delta T/T$.  The comparison between the simulations and the measurements is seen to be very good.  The model reproduces the two distinct time scales observed experimentally in the transmission recovery transients at low substrate temperatures.  The substrate temperature dependencies of the fitting parameters $\tau$ and $\tau_{opt}$ (Table.1) are in reasonable agreement with the expectations \cite{Kim10,KangOPLife10,Heinz11}.  However, the mechanism responsible for the temperature dependence of the doping density $N_{d}$, although small, is unclear.   

\begin{figure}[tbp]
	\centering
		\includegraphics[width=.40\textwidth]{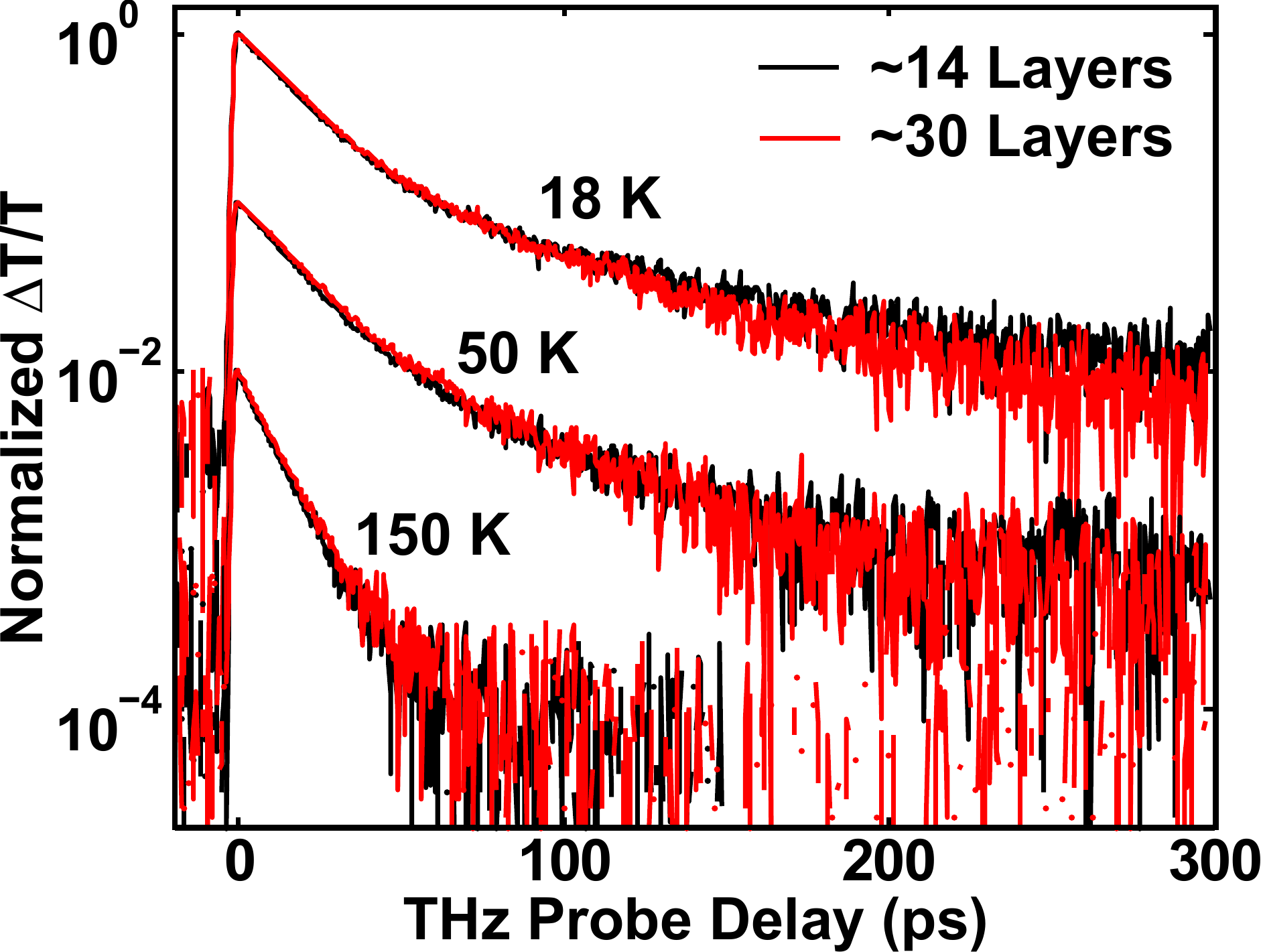}
	\caption{Measured $\Delta T/T$ transients for  $\sim$14 ML and $\sim$30 ML epitaxial graphene samples are plotted for different substrate temperatures.  $\Delta T/T$ transients for the 14 ML and 30 ML samples are normalized to the same peak magnitude and shifted along the vertical axis at each substrate temperature for clarity.  Although the sample thicknesses differ by a factor of $\sim$2, the measured relaxation time scales are nearly identical.}
	\label{fig:lowTcomp}
\end{figure}

In order to explore whether heat transfer either among the graphene layers or between the graphene layers and the substrate could be responsible for any of the features observed in our experiments \cite{Mak10}, we performed measurements on epitaxially grown graphene samples with different numbers of graphene layers.  However, we observed no significant changes in the observed time scales between the samples.  Fig.\ref{fig:lowTcomp} shows the results obtained for samples having 14 and 30 mono layers (ML) of graphene at different substrate temperatures.  For clarity in comparison, $\Delta T/T$ values shown in Fig.\ref{fig:lowTcomp} have been normalized so that the peak values for the 14 ML and 30 ML samples are equal.  The time scales associated with the transients are seen to be virtually identical despite the large difference in the number of graphene layers between the two samples.  These observations are consistent with our theoretical model where, as discussed earlier, the values of the time constant $\tau_{sub}$ in the 25-100 ps range are seen to have minimal effect on the simulation results.

To conclude, we have studied the relaxation dynamics of photoexcited carriers in epitaxial graphene at different substrate temperatures using optical-pump terahertz-probe spectroscopy.  The observed differential terahertz transmission transients show very long relaxation times extending out to hundreds of picoseconds at low substrate temperatures.  We also have presented a model that is in good agreement with the measurements and shows that carrier cooling in graphene can be very slow if the bulk of the carrier energy distribution falls low enough that both the intraband and interband optical phonon emission processes are not possible.

The authors acknowledge helpful discussions with Paul L. McEuen and Jiwoong Park, and acknowledge support from the National Science Foundation (monitor, Eric Johnson), the DARPA Young Faculty Award, the MURI program of the Air Force Office of Scientific Research (monitor, Harold Weinstock), the Office of Naval Research (monitor, Paul Makki), and the Cornell Material Science and Engineering Center (CCMR) program of the National Science Foundation.

\bibliography{OPTPGRbib}

\end{document}